\documentclass[conference]{IEEEtran}
\IEEEoverridecommandlockouts
\usepackage{amsmath,amssymb,amsfonts}
\usepackage{algorithm}
\usepackage{algpseudocode}
\usepackage{graphicx}
\usepackage{textcomp}
\usepackage{xcolor}
\usepackage[utf8]{inputenc}    %
\usepackage{balance}

\usepackage[defernumbers=true,bibstyle=ieee,citestyle=numeric-comp,bibencoding=utf8,backend=bibtex,style=ieee]{biblatex}
\begin{document}
\title{RSSI-based Outdoor Localization with \\Single Unmanned Aerial Vehicle\\
}
\author{\IEEEauthorblockN{Seyma Yucer\IEEEauthorrefmark{1}\IEEEauthorrefmark{3}, Furkan Tektas\IEEEauthorrefmark{1}, Mesih Veysi Kilinc\IEEEauthorrefmark{1}, Ilyas Kandemir\IEEEauthorrefmark{4}, \\
Hasari Celebi\IEEEauthorrefmark{1}\IEEEauthorrefmark{3}, Yakup Genc\IEEEauthorrefmark{3}, Yusuf Sinan Akgul\IEEEauthorrefmark{3}}

\IEEEauthorblockA{\textit{Institute of Information Technologies\IEEEauthorrefmark{1},Computer Engineering Department\IEEEauthorrefmark{3},Mechanical Engineering Department\IEEEauthorrefmark{4}}\\
\textit{Gebze Technical University}\\
Kocaeli, Turkey }
}

\maketitle
\begin{abstract}
Localization of a target object has been performed conventionally using multiple terrestrial reference nodes. This paradigm is recently shifted towards utilization of unmanned aerial vehicles (UAVs) for locating target objects. Since locating of a target using simultaneous multiple UAVs is costly and impractical, achieving this task by utilizing single UAV becomes desirable. Hence, in this paper, we propose an RSSI-based localization method that utilizes only a single UAV. The proposed approach  is based on clustering method along with the Singular Value Decomposition (SVD). The performance of the proposed method is verified by the experimental measurements collected by a UAV that we have designed and computer simulations. The results show that the proposed method can achieve location accuracy as low as 7m depending on the number of iterations.        
\end{abstract}


\section{Introduction}
In recent years, Unmanned Aircraft Systems (UAS) drew significant attention due to the proliferation of their usage in the places that are not easy to reach by humans \cite{Santamaria2007}. UAS play a critical role in many applications including search and rescue, public security and natural disaster aid \cite{Shi2018,Waharte2010}. 

One of the most critical parts of these systems is the positioning of targets and Unmanned Aerial Vehicles (UAV) itself \cite{Lazzari2017}. Although the GPS became a de-facto standard for outdoor positioning \cite{Sutheerakul2017}, especially for UAS, it can become obsolete in rural areas or in cases of natural disasters \cite{Rady2011} and GPS is not preferred due to the high energy consumption of GPS sensors. In such cases, RF signals can be exploited to be used as a positioning solutions \cite{Bittencourt2018,Dehghan2014} instead of GPS.

RF signal-based outdoor positioning methods use different properties of RF signals, for instance, Angle of Arrival, Time of Arrival, Time Difference of Arrival, and Received Signal Strength Indicator (RSSI). Unlike other methods, RSSI-based outdoor positioning methods \cite{Kuo2010,Dong2014} do not require a specialized hardware and much more affordable, since commodity hardware can provide RSSI of RF signals.

RF signal-based methods utilize  \cite{Oguejiofor2013,Yang2016} received signal strength indicator (RSSI) by employing the signal propagation model. Since the outdoor environment can have a dynamic and complex settings, signal propagation model \cite{Alippi2006} fails to satisfy positioning requirements. To overcome such obstacles, a study \cite{Naik2008} uses multiple RSSI, i.e., RSSI fingerprints, from various sources in order to estimate position by creating a radio map, an RSSI fingerprint model, of the environment. These passive \cite{Tang2016,Deak2011} methods consist of offline survey phase for creating the radio map and the online positioning phase. When comparing to trilateration-based approaches \cite{Oguejiofor2013}, most of the time radio map-based approaches can achieve more accurate and robust results. Data selection and preprocessing are the vital parts of radio map-based approaches since they are more sensitive to noise and environment dynamics.

Moreover, active positioning methods provide a position estimation without a priori exploration of the RF signal behavior in the environment. Therefore, these methods also need to confront the same challenges as the radio map-based methods do. These methods become more challenging when positioning occurs while moving. Localization in UAS is an excellent example of such cases as some UAVs can fly tens of meters each second. Most of the methods \cite{Khelifi2018,Dehghan2012} provide promising results by using simulation data, yet they would fail when applied to real-world scenarios.

In this paper, we propose an RSSI-based localization method which utilizes clustering method as a pre-processing module and Singular Value Decomposition (SVD) for positioning target transmitter node using only a single UAV instead of multiple UAVs. We present the performance of our method using a fixed-wing UAV in a real-world setup. In particular, our contributions are three-folds:
\begin{enumerate}
\item Requiring only a single UAV for localization instead of multiple UAVs,
\item A pre-processing method to eliminate less relevant or outlier RSSI samples in order to increase efficiency regarding the processing power and energy consumption while preserving real-time positioning capabilities
\item A hybrid model which combines the powerful features of active and passive positioning methods in order to provide robust location accuracy.
\end{enumerate}
This paper is organized as follows. Section \ref{sec:hardwaredesign} describes UAV's hardware design. Then, Section \ref{sec:proposedmethod} introduces the details of our proposed method. Section \ref{sec:experimentalresults} discusses the experimental results. Finally, the last section provides conclusions and directions for future work.

\section{Hardware System Design}\label{sec:hardwaredesign}
\subsection{The UAV Design}\label{GeneralSetup}
\begin{figure}[tp]
        \centerline{\includegraphics[width=\columnwidth]{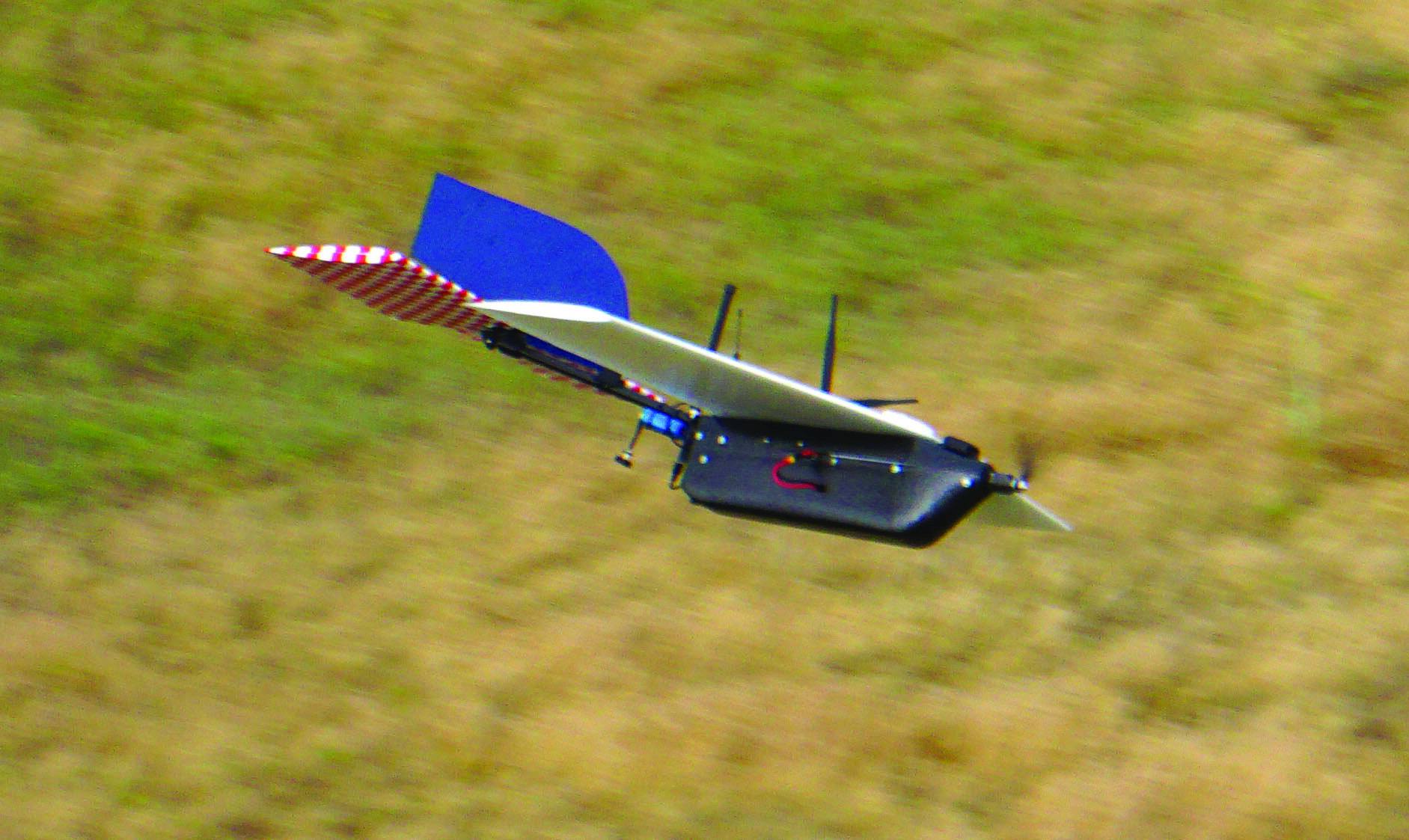}}
        \caption{Designed UAV performs an autonomous flight}
        \label{fig:uav-ota}
    \end{figure}

In order to conduct RSSI-based outdoor localization experiments for this paper, we designed a special fixed-wing UAV with minimum moving surfaces. This UAV constructed with a tractor configuration and has two independently-driven elevon surfaces. The designed UAV can be seen in Figure \ref{fig:uav-ota} while it is controlled via autopilot software. Hardware used in the UAV is listed in Table \ref{tab:uavhardwarecharacter}.
 
To conduct experiments of this paper, we launched predefined measurement campaigns in each flight. UAV performs these surveys in autonomous mode using Navio2, Autopilot card attached to Raspberry Pi 3. It is worthy to mention that Navio2 uses ArduPilot software which is a real-time flight control system that is empowered with MAVLink protocol. The UAV communicate with Ground Control Station (GCS) over a full-duplex telemetry module via the UART interface of the Navio2 with 57600 bps.

\begin{table}[hp]
\caption{Hardware Characterization of the UAV}
\begin{center}
\begin{tabular}{|c|c|}
\hline
\cline{1-2} 
Flight controller & Navio2 + Raspberry Pi 3 \\
	\hline
Weight & 2200g  \\
	\hline
Engine & NTM Prop Drive 28-36 1400KV \\
	\hline
Propeller & APC 9.0/4.5 \\
	\hline
Telemetry & RFD 900+ \\
	\hline
Battery & MultiStar LiPo 4S2P 8000mAh \\
	\hline
RC & Arkbird 433Mhz UHF 10-CH \\
	\hline

\end{tabular}
\end{center}
\label{tab:uavhardwarecharacter}
\end{table}

\begin{figure}[b]
        \centerline{\includegraphics[width=\columnwidth]{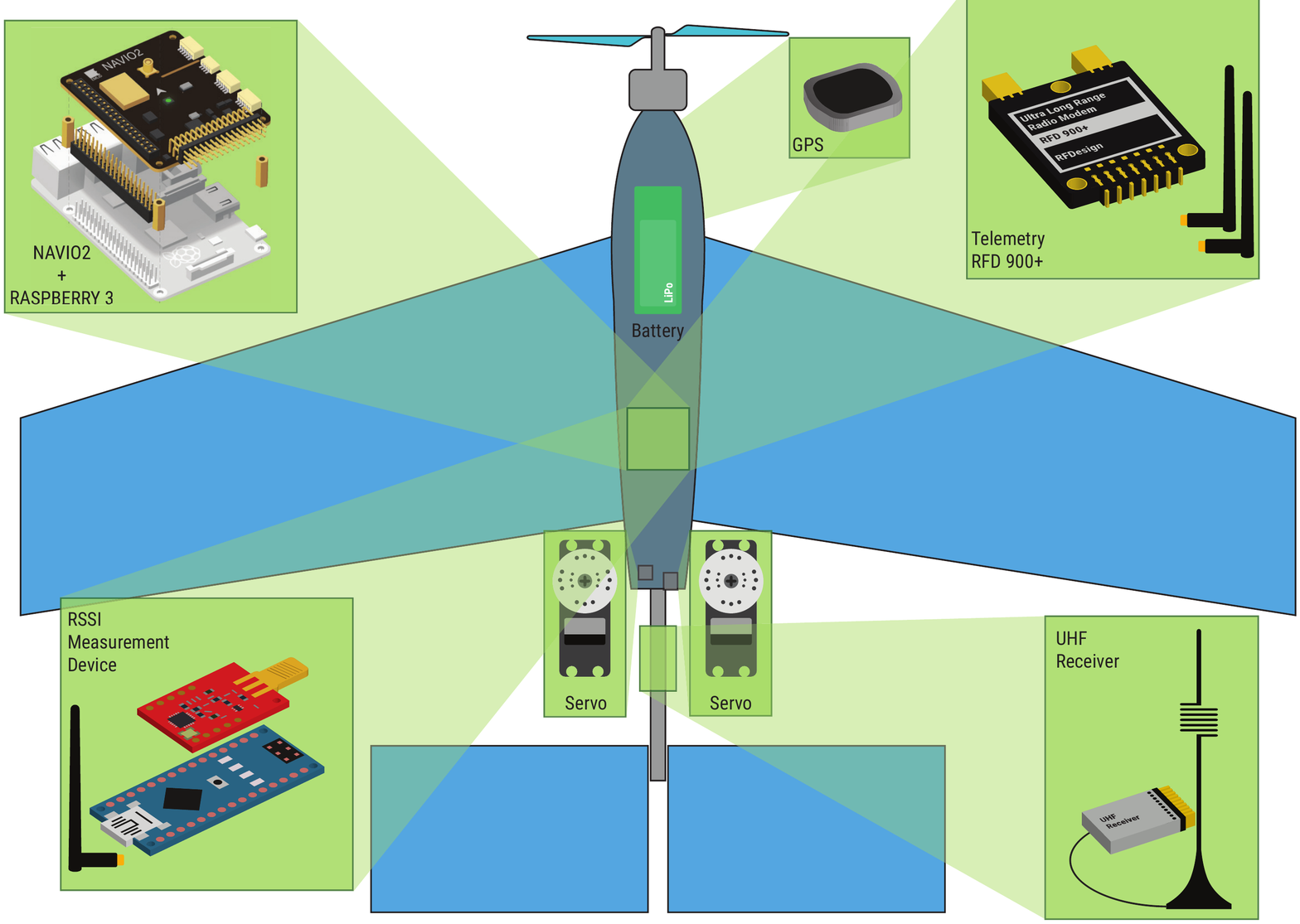}}
        \caption{UAV hardware configuration for RSS observation flights}
        \label{fig:uav-arch}
    \end{figure}

Each UAV is equipped with six antennas. The first antenna is used to receive GPS signals and attached to Navio2. Two of them are used for the different polarization in the telemetry module. The forth antenna used to gather RSSI samples from the target node. The fifth antenna emits an analog video stream to the ground with an On-screen Display telemetry information, and the last antenna is used for the remote controller. Note that the UAV hardware placement scheme can be seen in Figure \ref{fig:uav-arch}.

MAVLink is a communication protocol widely used between aerial vehicles and ground control stations as well as inside the aerial vehicles itself. MAVLink protocol can be extended to exchange custom message types in addition to its own predefined message types. We transferred RSSI measurements of the UAV to the GCS using a custom MAVLink message. These messages contain UAV's current GPS position and the RSSI sensor readings for different channels. 

RSSI module was calibrated with an external signal generator to report the same RSSI value from the same distance in different measurement campaings. UAV marshalls RSSI measurements with the current GPS position of the UAV, then transmits them to the GCS. After having certain number of RSSI observations, GCS estimates the target position using the collected RSSI readings.

\subsection{RSSI Measurement Instrument}\label{AA}

RSSI measurement device is used to gather RF signals of desired frequency with pre-defined bandwidth and then calculate the signal strength of the measured RF signal. This sensor is able to linearly measure -10 dBm -120 dBm band with 0.5 dBm steps. The measurement module was developed in-house. It consists of two sub-modules.
\begin{itemize}
    
    \item The first module is a RF transceiver, which uses SI4463 IC. It can measure the RSSI of RF band of interest which it has been tuned to within 12.5 KHz bandwidth. Transceiver uses SPI to communicate with microcontroller module.
    \item The second module is the Arduino Nano, a microcontroller module. It communicates with RF module, reads RSSI measurements, sorts them within a time window and sends the maximum reading to the onboard computer via UART communication bus. A custom communication protocol is designed between onboard computer and micro controller module. The microcontroller module gathers RSSI measurements from RF transceiver and sends them to onboard computer at 1Hz.
\end{itemize}

\section{The Proposed Method} \label{sec:proposedmethod}

In this paper, our aim is to localize an active stationary RF transmitter using a single fixed-wing UAV. Since having multiple UAVs on the air is not always possible and feasible, we try to achieve robust positioning accuracy with a single UAV by eliminating the erroneous RSSI measurements and selecting the most accurate ones. In the following sections, we assume that UAV has a dedicated GPS module along with a RF receiver module to transmit RSSI measurements and GPS coordinates to the GCS over a telemetry link. We will explain our method in three subsections: (a) RF pathloss model, (b) proposed multiple node clustering method, and (c) multilateration model for UAVs.  

\subsection{RSSI Path-loss Model}

Active localization techniques require multiple reference nodes to be available simultaneously in the air to estimate the target position with respect to these reference nodes. The RSSI values obtained from these reference nodes are converted to a distance metric, individually. Thus, these techniques benefit from the Friis transmission formula (\ref{eq:friis}), \cite{Friis1946} as we also use in our proposed method.

\begin{equation}
\label{eq:friis}
P_r = P_t  G_t  G_r  \left(\frac{\lambda}{4 \pi d}\right)^2
\end{equation}

In (\ref{eq:friis}), \(P_r\) is the received power, \(P_t\) is the transmission power, and ${\lambda}$ is the wavelength of the signal. \(G_t\) and \(G_r\) denote the antenna gain of the transmitter and receiver, respectively. The term \(d\) is the relative distance between transmitter and receiver devices. The relation between RSSI values and distance is defined in this formula.

The received RSSI fingerprints are transmitted to the GCS and converted into relative real world distances using the log distance path loss formula \cite{Rappaport2002}:

\begin{equation}
\label{eq:2}
d = d_010^\frac{-P_\textit{r}(d)+P_\textit{0}(d_\textit{0})- {\textit{X}}}{10n}
\end{equation}

\begin{figure}[ht]
        \centerline{\includegraphics[width=\columnwidth]{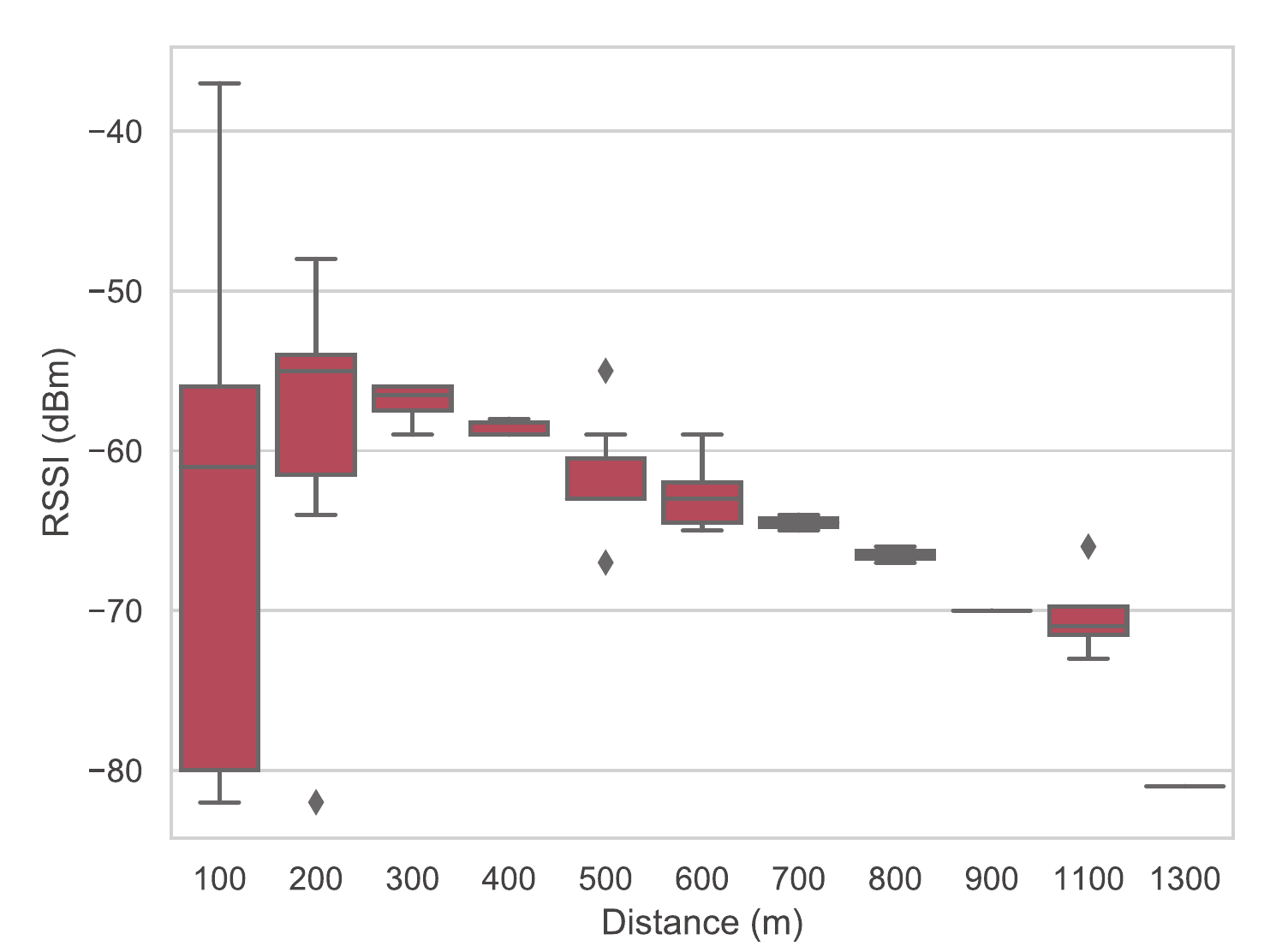}}
        \caption{The noise effect on the RSSI measurements at different distances}
        \label{fig:rssidbmdist}
    \end{figure}

The value of the relative distance, \(d\), can be calculated using (\ref{eq:2}) where \(\textit{X}\) is a Gaussian distributed random variable. In order to estimate \(d\), we need to determine \(n\), signal propagation constant, \(d_0\), the reference distance value, beforehand and measure the \(P_0(d_0)\), the reference power value in \textit{dBm} at \(d_0\). It is worthy to note that, weather conditions and environment dynamics have a dramatic effect on these measurements. So that, we measured these reference values before each survey during the first loiter of the flight. These reference values are also affected by noise which has a dramatic effect on localization accuracy. To show the noise effect on RSSI measurements at the same distance, we visualized the relationship between RSSI values and the distance where we have multiple measurements at the same distance in the Figure \ref{fig:rssidbmdist}. Therefore, it is important to select the most reliable RSSI measurements for accurate positioning. In the next section, we will elaborate on our proposed RSSI selection approach.

\subsection{Multiple Node Clustering}

In our experiments, we observed that the fluctuations of RSSI values needs to be addressed to achieve a robust localization accuracy. In order to do so, we propose a dynamic clustering module which eliminates the noisy RSSI measurements as well as applies a threshold to RSSI values below a certain \textit{dBm} value. This module also selects the most eligible RSSI measurements among neighbor RSSI measurements. We observed that this selection concludes better positioning accuracy over using all available RSSI measurements.

There are several methods for selecting the most representative RSSI measurements such as grid planning and hierarchical clustering. For instance, grid planning method slices the survey area into a rectangular grid then selects either minimum, average or maximum RSSI value within each rectangular area. Our module dynamically clusters the latitude and longitude coordinates of the given reference node. Then we obtain the cluster centers and the cluster membership for every RSSI measurement.

In (\ref{eq:3}), where $i,j \in \{0,1,2, \dots, n\}$, we define a simple reference node which has the latitude coordinate of \(x_i\) and longitude coordinate of \(y_i\). To cluster these reference nodes, we preferred K-means clustering, an efficient and popular unsupervised clustering technique. In (\ref{eq:4}), we select \(\kappa\) value depending on the \(m_a\), an empirical constant distance, and the maximum distance between given reference nodes. \(H(p_i,p_j)\) in (\ref{eq:4}) represents the Haversine distance formula \cite{H.1930} for calculating the real-world distance using the latitude and longitude coordinates. 
\begin{equation}
\textit{\textbf{\(p_i\)}}  =
\begin{bmatrix}
x_i \\
y_i 
\end{bmatrix}
\label{eq:3}
\end{equation}

\begin{equation}
\kappa =\frac{max(\textbf{\(H\)} (p_i, p_j))}{ m_a}
\label{eq:4}
\end{equation}

Our method groups each RSSI measurement into a geographically defined cluster, then selects the most powerful RSSI values within each cluster, denoted as \({P_n}^{max}\). It also eliminates the clusters which contain less number of RSSI measurements. Then using the most powerful RSSI values, we convert this problem into a linear form and solve it using a Singular Value Decomposition (SVD) technique. We will describe how we dynamically applied this method in the following section.

\subsection{Multilateration Model for UAVs}

The multilateration technique consists of matrix-based simple operations and the accuracy of this technique strongly depends on minimum three RSSI measurements used in creation of this matrix.

In UAVs, RSSI measurements deviate due to several factors including vehicle speed, wind, trees and buildings and so on. Therefore, we need a positioning method that rely on multiple representative measurements. Assume that we have M reference measurements including latitude and longitude, and \(d_i\) is the relative distance between target node and the reference node at \(P_i\). First, we need to transform latitude \(x_i\) and longitude \(y_i\) to the projection coordinates, \(x_i'\), \(y_i'\), respectively. Then, we use (\ref{eq:relationdandxy}) from \cite{Yang2016} to find the target node coordinates by solving the following equations. We can form (\ref{eq:axb}) using (\ref{eq:relationdandxy}) where $A$ is an \textit{M x N} matrix, \textrm{x} is an \textit{N x 1} matrix, and $b$ is an \textit{M x 1} matrix.

Matrix $A$ contains the reference measurement coordinates and matrix $b$ contains distances between target node and reference node within the same row. Therefore, using the coordinates and distances between nodes, we can transform the multilateration localization problem into a linear problem, so that we can estimate the position of the target node by solving (\ref{eq:axb}).

\begin{equation}
\label{eq:relationdandxy}
\begin{split}
(x'-x_1')^2 + (y'-y_1')^2  &= d_1^2,\\
(x'-x_2')^2 + (y'-y_2')^2  &= d_2^2,\\
&.\\
&.\\
(x'-x_m')^2 + (y'-y_m')^2  &= d_m^2.
\end{split}
\end{equation}

\begin{equation}
\label{eq:axb}
A\textrm{x'} = b
\end{equation}

\begin{equation}
\begin{split}
\label{eq:linearmatrixs}
\textit{\textbf{\(A\)}}  =&
\begin{bmatrix}
1 & -2x_1' & 2y_1' \\
1 & -2x_2' &2y_2'  \\
. & . &  .      \\
. & . & .       \\
1 & -2x_m' & 2y_m' \\
\end{bmatrix}
\textrm{x'} =
\begin{bmatrix}
x'^2 + y'^2       \\
x'               \\
y'               \\
\end{bmatrix}
\\
\textit{\textbf{\(b\)}}  =&
\begin{bmatrix}
d_1^2 - x_1'^2 - y_1'^2 \\
d_2^2 - x_2'^2 - y_2'^2 \\
.                     \\
.                      \\
d_m^2 - x_m'^2 - y_m'^2 \\
\end{bmatrix}
.
\end{split}
\end{equation}

After solving the linear equation using SVD, we transform \textrm{x'} matrix' elements into the WGS84 real world coordinate system to obtain geographic coordinates of the target node, \(x\) and \(y\).

\begin{algorithm}[htbp]
\renewcommand{\thealgorithm}{}
\caption{An iteration of the proposed localization algorithm}
\begin{algorithmic}
\renewcommand{\algorithmicrequire}{\textbf{Input:}}
\renewcommand{\algorithmicensure}{\textbf{Output:}}
\Require $R$: list of all prior RSS observations
\Require $G$: list of all prior GPS coordinates
\Ensure  Target coordinates and iteration error
\Function{estimate\_target\_position}{$R,G$}
    \For{$i \in R$}
        \For{$j \in R$}
            \State $d[i][j] \gets \left\lVert R_i,R_j \right\rVert $
        \EndFor
    \EndFor
    \State $d_{max} \gets \Call{max}{d}$
    \State $K \gets d_{max} / m_a$
    \State $RF \gets (R, G)$
    \State $C \gets \Call{KMeans}{K, RF}$
    \State $C^\prime \gets list() $
    \ForAll{$c$ in $C$}
        \If{$c > R_{thresh}$} \Comment{$R_{thresh}$ is a constant}
        \State $C^\prime.append(c)$ 
        \EndIf
    \EndFor
    \State ${A,b,\textrm{x'}} \gets \Call{Multilateration}{C^\prime, d}$
    \State $G_{target} \gets \Call{WGS84-Transform}{\textrm{x'}} $
    \State ${error_{iteration}} \gets A \cdot \textrm{x'} - b$
    \State \Return ${G_{target}, error_{iteration}}$
\EndFunction
\end{algorithmic} 
\end{algorithm}

During flights, we measure RSSI values at constant intervals. Therefore, in order to cope with continuous flow of RSSI measurements, we need to design an iterative localization system which improves itself over time. In every iteration, we gather a certain number of RSSI measurements, then estimate the target position using the given algorithm. In the first step of the algorithm, we group all available measurements into $\kappa$ clusters and eliminate ineffective clusters. Then we find the most powerful RSSI measurement within each cluster and use them as reference nodes. We solve the linear system described above using these reference nodes, and obtain target coordinate estimation for the current iteration. We calculate linear model error for the current iteration using this estimation. Then we estimate the final target coordinate using the iteration with minimum iteration error among all previous iterations.
 
\section{Experimental Results}\label{sec:experimentalresults}
We evaluate our method using real flight measurements as well as simulation flight measurements. In both cases, we use the same evaluation criteria and report the error between target position and estimated position using the haversine distance.

To create a simulation environment, we created a fixed-wing UAV simulator using ArduPilot's test software. Using this UAV, we simulated a real measurement campaign inside the GTU campus and collected RSSI measurements at every second using (\ref{eq:friis}) for each GPS coordinate generated by the simulated UAV.

 \begin{figure}[ht]
    \centerline{\includegraphics[width=\columnwidth]{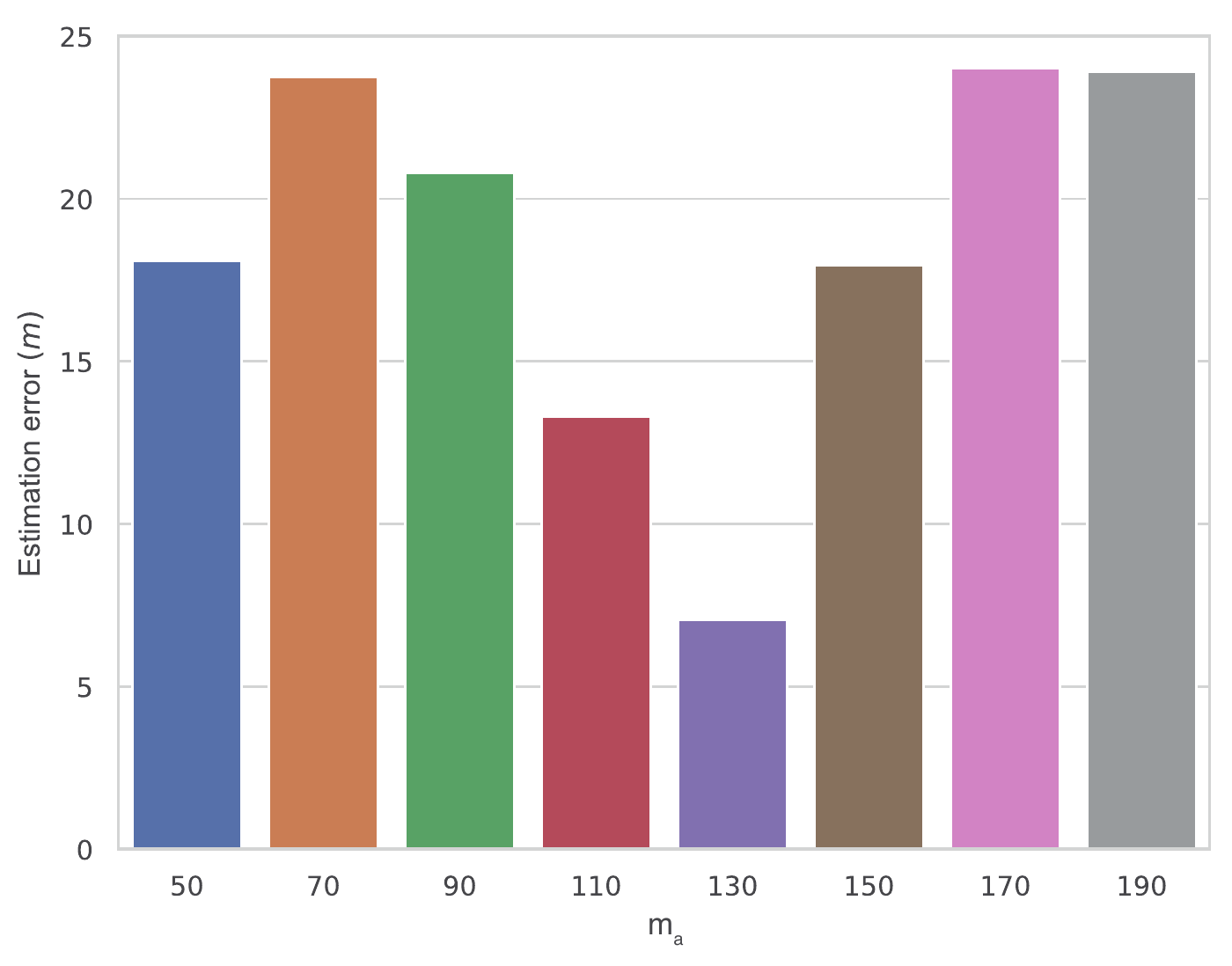}}
    \caption{Estimation error in meters for different $m_a$ values on the simulation data}
    \label{fig:ma_vals}
\end{figure}

 \begin{figure}[hb]
    \centerline{\includegraphics[width=\columnwidth]{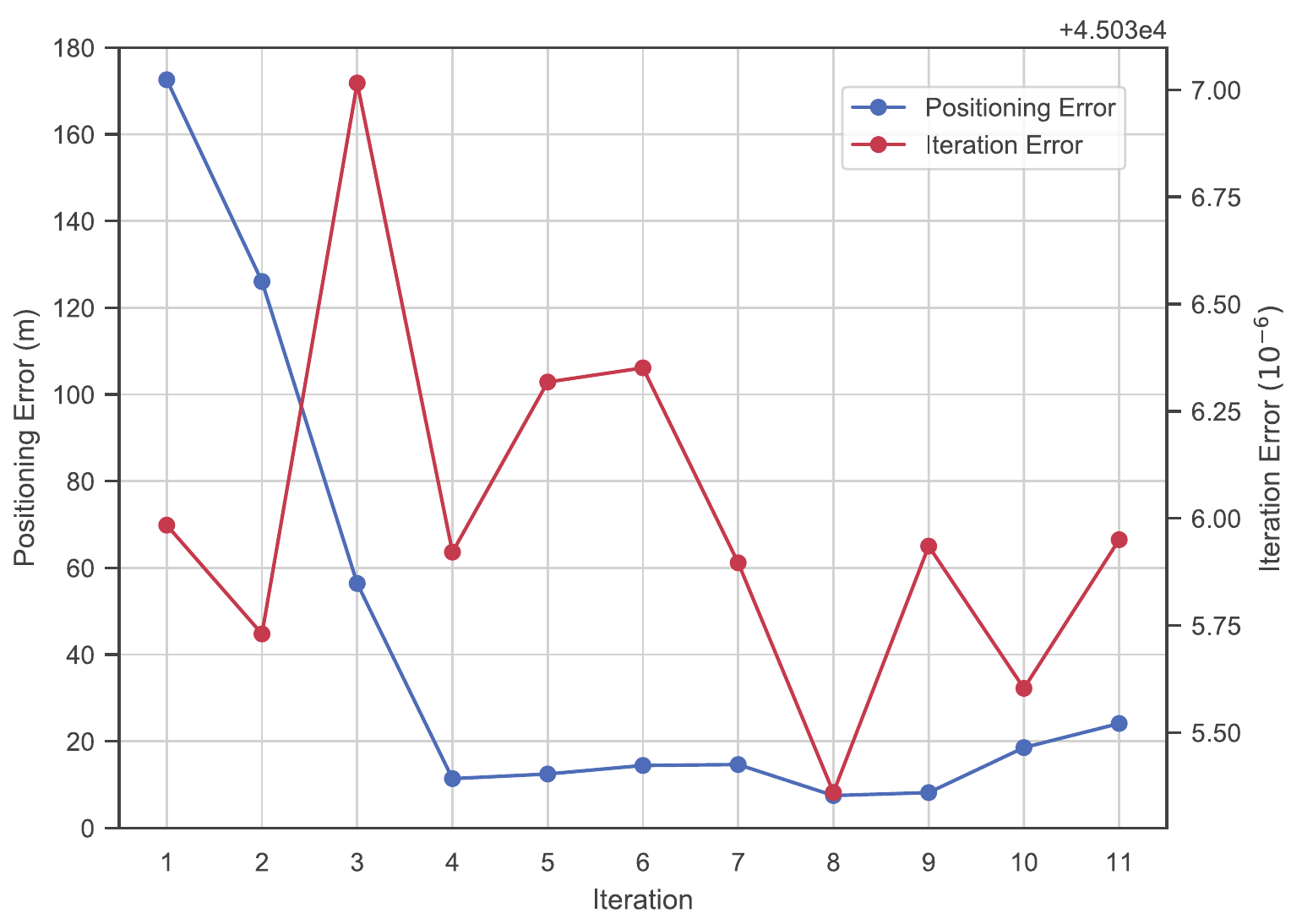}}
    \caption{Estimation and iteration errors for $m_a = 130$ on the simulation data}
    \label{fig:ma_130_est_err}
\end{figure}

The $m_a$ value has a strong impact on the position estimation error. Hence, it is important to select appropriate $m_a$ value to mitigate estimation errors. We tested and plotted the effect of different $m_a$ values ranging from 50 to 190 on estimation errors in Figure \ref{fig:ma_vals}. Using our simulation and field observations, we empirically select ${m_a}$ value as 130 for simulation, 50 for real world surveys and $R_{thresh}$ equals to iteration number.  For every \(50^{th}\) RSSI measurement (i.e. iteration), we estimate the target position and $error_{estimation}$. Simulation data position estimation errors when $m_a = 130$ can be seen in Figure \ref{fig:ma_130_est_err} where \(8^{th}\) iteration has the minimum iteration error which results in 7 meter position estimation error.

In addition to simulation results, we tested our proposed method in the field using a fixed-wing UAV described in the previous section. The flight measurement campaigns are conducted in different weather conditions and climates in GTU campus and $3^{rd}$ Air Wing in Konya. Before each flight, we measure the reference RSSI value and the signal path-loss exponent, \(n\), with a $100m$ distance inside the campaign area using a similar hardware setup of the UAV's. We calculated the \(n\) constant using our previous observations. These two survey areas differ from each other in terms of size and Line-of-Sight (LOS) availability. In Konya, survey area has a clear LOS for more than 7\(km\) whereas GTU campus have at most 150\(m\) clear LOS. In Table \ref{tab:all_results} we present the position estimation errors in meters as a haversine difference between target and estimated positions. SVD denotes the plain SVD method which takes all available measurements into account.

\begin{figure}[tbp]
    \centerline{\includegraphics[width=\columnwidth]{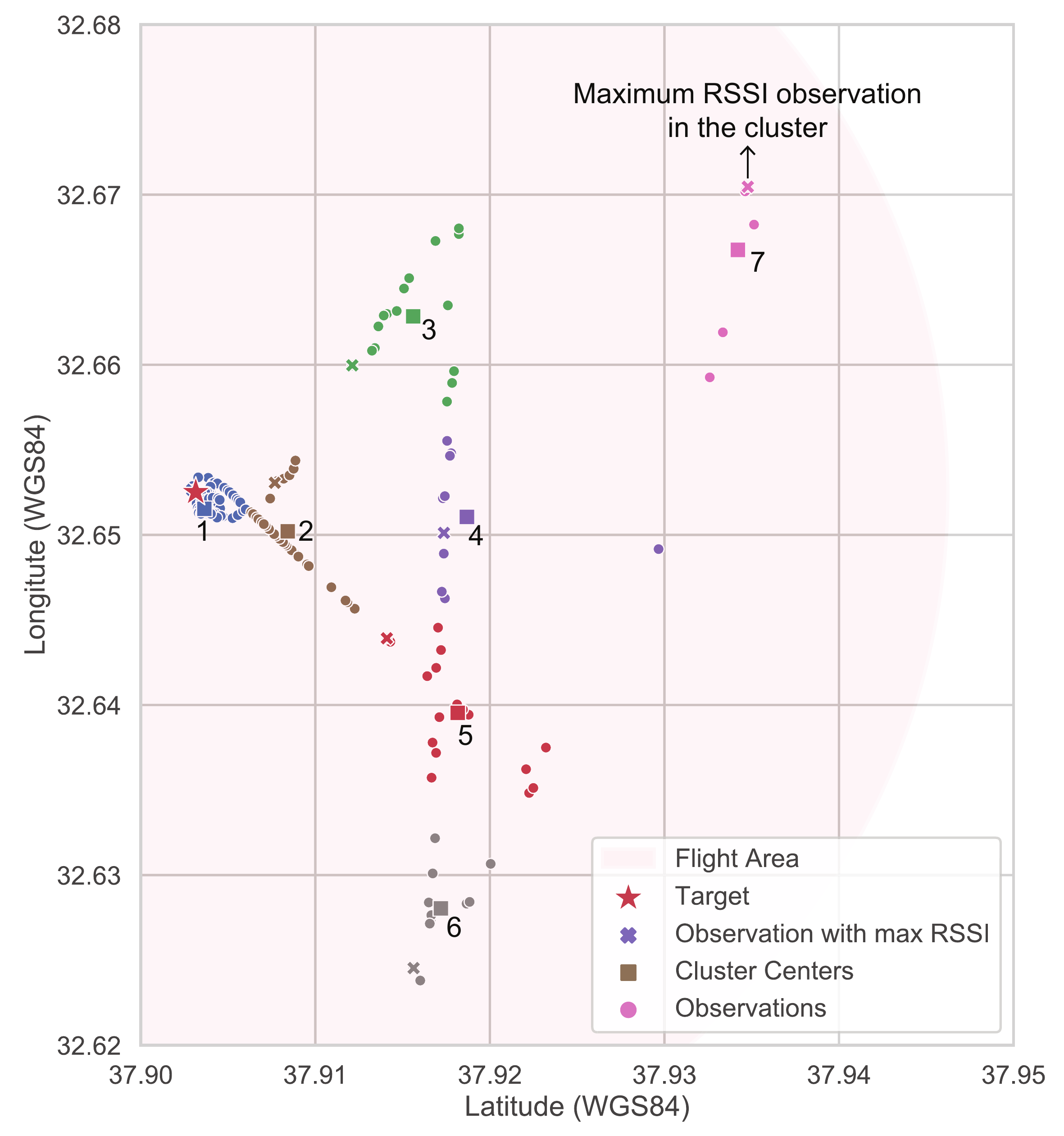}}
    \caption{Cluster visualization of the \textit{Konya flight}}
    \label{fig:konya_cluster_vis}
\end{figure}

In Figure \ref{fig:konya_cluster_vis}, we illustrate how our method works in real-life conditions and
the flight area centered in the target node is represented with the blush color. As can be seen in the figure, our method employs the spatial information of RSSI measurements, therefore it's more robust than plain SVD method since our method uses only the reference points and diminishes the effect of outlier measurements or eliminates them, entirely. 

\begin{table}[htbp]
\centering
\caption{Error in meters for estimating target location using Our Method and SVD.}
\begin{tabular}{cc|c|c|}
\cline{3-4}
\multicolumn{1}{l}{}                   & \multicolumn{1}{l|}{} & \multicolumn{2}{c|}{Estimation Error} \\ \hline
\multicolumn{1}{|c|}{Data}             & Survey Area ($km^2$)     & SVD ($m$)            & Our Method ($m$)           \\ \hline
\multicolumn{1}{|c|}{GTU - Simulation} & 3.14                   & 23                 & 7                 \\ \hline
\multicolumn{1}{|c|}{GTU - Flight}     & 0.95                  & 157                 & 43                 \\ \hline
\multicolumn{1}{|c|}{Konya - Flight}   & 70.51                 & 435                & 145                \\ \hline
\end{tabular}
\label{tab:all_results}
\end{table}

\section*{Conclusions}\label{sec:conclusions}
In this paper, we propose a RSSI-based localization method using only a single UAV instead of multiple UAVs to locate a target object. In the proposed method, we have applied clustering along with an SVD on RSSI measurements for positioning a target. In addition, a fixed-wing UAV is designed in order to demonstrate the performance of the proposed method. The flight measurements using this fixed-wing UAV and computer simulations are conducted. The results are promising for our proposed method that can be utilized for commercial and surveillance applications such as search and rescue. As a future work, the proposed method can be integrated and utilized in 5G networks.

\section*{Acknowledgment}
This study is supported by Turkish Air Force. We would like to thank Mete Can Gazi, Halis Kilic, Fatih Fahreddin Ongul, and Ahmet Soyyigit for their help during the measurement campaigns.
\balance
\printbibliography

\end{document}